\documentclass[aps,prl,twocolumn,superscriptaddress]{revtex4-1}

\usepackage{graphicx}
\usepackage{dcolumn}
\usepackage{bm}
\usepackage{amssymb}
\usepackage{latexsym}

\begin{document}


\title{Observation of Anderson localization in ultrathin films of three-dimensional topological insulators}
\author{Jian~Liao,$^1$ Yunbo~Ou}
\affiliation{Beijing National Laboratory for Condensed Matter Physics, Institute of Physics, Chinese Academy of Sciences, Beijing 100190, China}

\author{Xiao~Feng}
\affiliation{State Key Laboratory of Low Dimensional Quantum Physics, Department of Physics, Tsinghua University, Beijing 100084, China}

\author{Shuo~Yang,$^1$ Chaojing~Lin,$^1$ Wenmin~Yang,$^1$ Kehui~Wu}
\affiliation{Beijing National Laboratory for Condensed Matter Physics, Institute of Physics, Chinese Academy of Sciences, Beijing 100190, China}

\author{Ke He}
\email{kehe@tsinghua.edu.cn}
\affiliation{State Key Laboratory of Low Dimensional Quantum Physics, Department of Physics, Tsinghua University, Beijing 100084, China}

\author{Xucun~Ma,$^2$ Qi-Kun~Xue}
\affiliation{State Key Laboratory of Low Dimensional Quantum Physics, Department of Physics, Tsinghua University, Beijing 100084, China}

\author{Yongqing Li}
\email{yqli@iphy.ac.cn}
\affiliation{Beijing National Laboratory for Condensed Matter Physics, Institute of Physics, Chinese Academy of Sciences, Beijing 100190, China}

\date{\today}

\begin{abstract}
 Anderson localization, the absence of diffusive transport in disordered systems, has been manifested as hopping transport in numerous electronic systems, whereas in recently discovered topological insulators it has not been directly observed. Here we report experimental demonstration of a crossover from diffusive transport in the weak antilocalization regime to variable range hopping transport in the Anderson localization regime with ultrathin (Bi$_{1-x}$Sb$_x$)$_2$Te$_3$ films. As disorder becomes stronger, negative magnetoconductivity due to the weak antilocalization is gradually suppressed, and eventually positive magnetoconductivity emerges when the electron system becomes strongly localized. This works reveals the critical role of disorder in the quantum transport properties of ultrathin topological insulator films, in which theories have predicted rich physics related to topological phase transitions.
\end{abstract}

\maketitle

The concept of Anderson localization has profoundly influenced our understanding of electron conductivity~\cite{Abrahams10}. While examples for disorder driven metal-insulator transition are abundant in three-dimensions (3D), the question of whether Anderson transitions exist in 2D has posed a lot of theoretical and experimental challenges. Scaling theory proposed by Abrahams et al. predicts that there are no truly metallic states in non-interacting 2D electron systems~\cite{Abrahams79,Lee85}. It was, however, discovered later that extended electron states may exist when electron-electron (e-e) interaction, spin-orbit coupling (SOC) or magnetic field comes in to play~\cite{Evers08,Spivak10}. The 3D topological insulators (TIs) discovered in recent years~\cite{Hasan10,Qi11} provide novel types of 2D electron systems that are of particular interest for study of the localization-delocalization problem.  The Dirac surface states of 3D TIs are believed to be topologically protected from localization due to its special symmetry class~\cite{Konig12,Fu12,Ostrovsky10,Ostrovsky07,Ryu07}.  Moreover, when a 3D TI thin film is sufficiently thin, the hybridization between the top and the bottom surface states opens an energy gap near the Dirac point~\cite{ZhangY10}, and it is suggested theoretically that the hybridization gap would drive the electron system to topologically different phase, such as a quantum spin Hall insulator or a trivial band insulator~\cite{Linder09,LiuC10,Lu10}. Even though a lot of work has been carried out on electron transport properties of TI thin films~\cite{Culcer12,Ando13,Bardarson13,YangWM13}, the fate of such electron systems under the condition of strong disorder (or in other words, whether Anderson localization could take place), still remains unclear.

In this work, we have studied electron transport in a large number of highly gate-tunable TI thin films with various thicknesses and chemical compositions. We found that only in ultrathin TI films in which surface hybridization and disorder effects are significant, hopping transport, a hallmark of strong localization~\cite{Mott79,Shklovskii91,Ovadyahu99}, can be observed. The observed temperature and magnetic field dependences of conductivity suggest that electron transport can be driven from the diffusive transport governed by weak antilocalization (WAL) and e-e interactions to the Mott-type variable range hopping in the strong localization (i.e.\ Anderson localization) regime. The capability of tuning the surface states across a wide range of conductivity provides valuable insight into the nature of electron transport in the TI thin films.

\begin{figure}
\centering
\includegraphics*[width=8.5 cm]{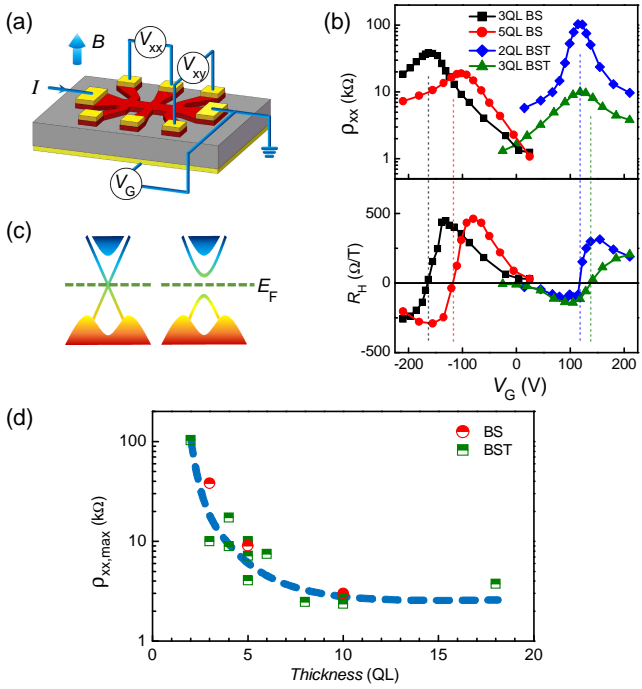}
\caption{(color online) (a) Sketch of a back-gated Hall bar device used for the electron transport measurements. The width of the current path is at least 50\,$\mu$m. (b) Gate voltage dependences of longitudinal resistance per square ($\rho_\mathrm{xx}$, top panel) and Hall coefficient (R$_H$, bottom panel) of four ultrathin TI films measured at $T$=1.6\,K. BST and BS denote Bi$_{1-x}$Sb$_x$)$_2$Te$_3$ and Bi$_2$Se$_3$, respectively. The sign of $R_H$ is set to positive for electrons. (c) Schematic band diagrams of the 3D TI thin films without (left) and with (right) the surface hybridization gap. (d)$\rho_\mathrm{xx,max}$ versus film thickness for 15 BST or BS samples. The dashed line shows the qualitative tendency of the thickness dependence.} \label{Fig1}
\end{figure}

The electron transport measurements were carried out on Hall bar structures of (Bi$_{1-x}$Sb$_x$)$_2$Te$_3$ and Bi$_2$Se$_3$ thin films (Fig.\,1a). For the (Bi$_{1-x}$Sb$_x$)$_2$Te$_3$ thin films with proper Bi:Sb ratio, ambipolar transport can be realized on a regular basis, owing to the low bulk carrier density in (Bi$_{1-x}$Sb$_x$)$_2$Te$_3$~\cite{ZhangJ11,Kong11} as well as the large gate tunability offered by the SrTiO$_3$ substrates~\cite{Chen10,He12}. Effectively tuning the chemical potential in Bi$_2$Se$_3$ is more challenging due to the high excess electron densities (on the order of 10$^{13}$ cm$^{-2}$) brought by Se vacancies and other sources~\cite{Analytis10}.
Fig.\,1b shows two examples of gate voltage tuning for (Bi$_{1-x}$Sb$_x$)$_2$Te$_3$ thin films and two for Bi$_2$Se$_3$. For each sample, the maximum in the longitudinal resistance per square, $\rho_{\mathrm{xx,max}}$, is located quite close to the gate voltage at which the Hall coefficient $R_H$ reverses its sign. This can be attributed to low sheet carrier densities of the bulk layer in the ultrathin TI films. The $\rho_\mathrm{xx}$ maximum can thus be regarded approximately as the charge neutral point (CNP)~\cite{Kim12}. In the region near the CNP (usually located between the minimum and the maximum of $R_H$), both electrons and holes exist in the film. Outside this region, the transport is dominated by either holes (left, smaller $V_G$) or electrons (right, larger $V_G$), and the responses to gating are similar to those of p- or n-type semiconductors.

 Fig.\,1c schematically illustrates the change of surface states from gapless in thicker films to gapped in ultrathin films. According to a photoemission study of Bi$_2$Se$_3$ thin films, the hybridization between the top and bottom surfaces opens a noticeable energy gap when thickness $t$ is reduced to 5 quintuple layers (5\,QL, $\approx$5\,nm) or less, and the gap can be as large as 0.25\,eV~\cite{ZhangY10}. This can qualitatively explain the data shown in Fig.\,1d, which summarizes the $\rho_\mathrm{xx,max}$ data from more than a dozen highly tunable (Bi$_{1-x}$Sb$_x$)$_2$Te$_3$ and Bi$_2$Se$_3$ samples with various thicknesses. For the films thicker than 6\,QL, $\rho_\mathrm{xx,max}$ is about 3\,k$\Omega$, nearly independent of the thickness. In contrast, $\rho_\mathrm{xx,max}$ for $t<6$\,QL increases rapidly with decreasing thickness. The small $\rho_\mathrm{xx,max}$ values obtained for thicker films are related to the gapless surface states, whereas the much larger $\rho_\mathrm{xx,max}$ observed in the thinner films can be attributed to the hybridization gap as well as stronger disorder effects.

\begin{figure}
\centering
\includegraphics*[width=8.5 cm]{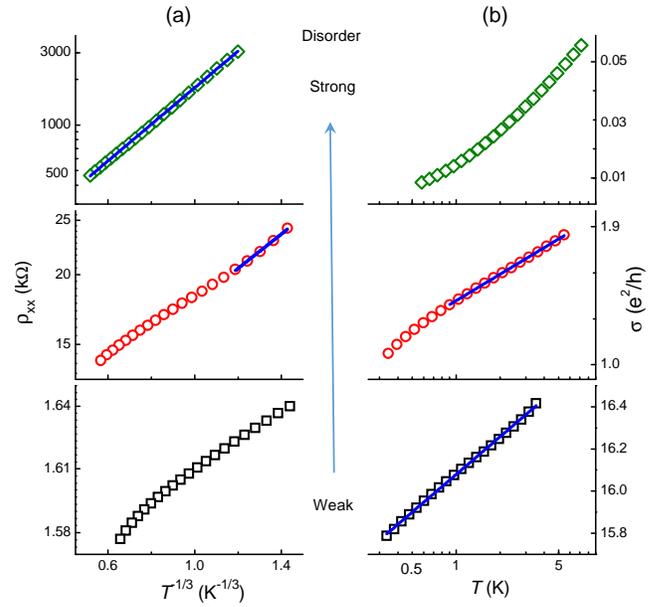}
\caption{(color online) Temperature dependences of $\rho_\mathrm{xx}$ (in logarithmic scale, panel a) and $\sigma$ (in linear scale, panel b), which are plotted as functions of $T^{-1/3}$ and $T$, respectively. As the electron system becomes more disordered (from bottom to top), the transport varies from weak antilocalization to the Mott-type variable range hopping.}
\label{Fig2}
\end{figure}

By combining precise thickness control with the tuning of chemical composition and gate voltage, we were able to obtain
$\rho_\mathrm{xx}$ values much larger than \,$h/e^2$ ($\approx25.8$\,k$\Omega$) in the films with $t\leq3$\,QL. The $\rho_\mathrm{xx}$ value can provide a convenient measure of how disordered the electron system is. The resistance $h/e^2$ is corresponding to $g\equiv\sigma/(e^2/h)=k_F l=1$ in the Drude model, where $g$ is the dimensionless conductivity with $\sigma=1/\rho_\mathrm{xx}$, $k_F$ is the Fermi wavevector, and $l$ is the mean free path. Such a resistance defines the Ioffe-Regel criterion, which states that $g>1$ needs to be satisfied in a conductor~\cite{Mott79}. The thickness dependence of $\rho_\mathrm{xx,max}$ shown in Fig.\,1d suggests that the transport can be gradually tuned from the diffusive regime ($g\gg1$)~\cite{Bergmann84} toward the strong localization regime ($g\ll 1$, corresponding to $\rho_\mathrm{xx}$ on the order of 100\,k$\Omega$ or higher)~\cite{Mott79,Shklovskii91}.

The change in the transport regime with increasing disorder is further manifested in the temperature dependence of conductivity depicted in Fig.\,2. For large conductivity ($g\gg$1), logarithmic dependence on temperature $T$ was observed, consistent with previous measurements of TI thin films in the diffusive regime~\cite{Chen11,LiuM11,Wang11,Takagaki12,Chiu13,Roy13}. For the TI surface states, the $\pi$ Berry phase associated with the spin-momentum locking leads to destructive interference for electron backscatterings and hence WAL, an effect opposite to the weak localization (WL) in ordinary 2D electron systems with negligible SOC. One would thus expect a $\ln T$ type of \emph{increase} in conductivity as $T$ drops. The e-e interactions~\cite{Altshuler85}, however, produce a larger, opposite $\ln T$ correction to the conductivity. The combined effects of WAL and e-e interactions can account for the observed $T$-dependence in the weakly disordered regime~\cite{Ostrovsky10,LuHZ14}. This is in contrast with $g\ll1$, at which the transport was found to follow Mott's law for the variable range hopping in 2D~\cite{Mott79,Ovadyahu99}, namely $\rho_\mathrm{xx} \propto \exp(T^{-1/3})$. For intermediate conductivity ($g\sim1$), there is a crossover from the logarithmic temperature dependence at higher $T$ to an exponential one at lower $T$. This can be understood as a consequence of competition between dephasing length $l_\phi=(D\tau_\phi)^{1/2}$ and localization length $\xi$. An electron system can be regarded as strongly localized when $l_\phi\gg\xi$ and weakly (anti)localized when $l_\phi\ll\xi$.  At sufficiently low temperatures, the dephasing length is expected to follow $l_\phi \propto T^{-1/2}$ since e-e interactions become the dominant source of dephasing~\cite{Altshuler85,Lee12,Cha12}. Lowering temperature makes $l_\phi$ exceeding $\xi$, and hence drives the crossover from WAL to strong localization. Similar transition can also be induced by varying the magnitude of disorder (see Fig.\,S6 in the online supplemental material~\cite{Supplem}).

\begin{figure}
\centering
\includegraphics*[width=8.5 cm]{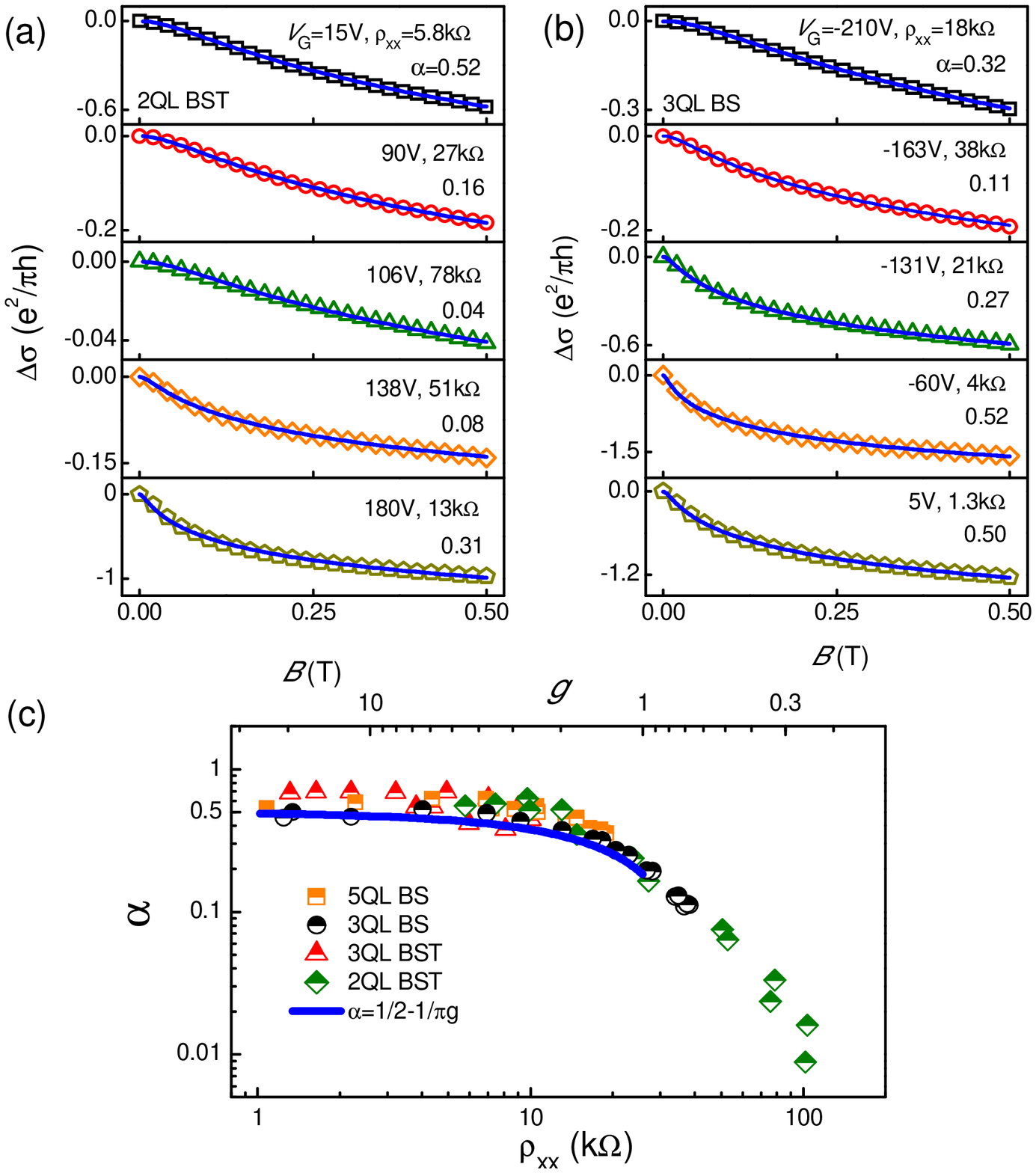}
\caption{(color online) (a-b) Representative low field magnetoconductivity curves and their best fits to the Hikami-Larkin-Nagaoka equation for two TI thin films: (a) 2\,QL thick Bi$_{1-x}$Sb$_x$)$_2$Te$_3$ (BST) thin film, (b) 3\,QL Bi$_2$Se$_3$ (BS) thin film. (c) Prefactor $\alpha$ extracted from the HLN fit plotted as a function of $\rho_\mathrm{xx}$ for four ultrathin TI films. The data extracted from the experiment are drawn in symbols, and the line is a theoretical curve that takes the disorder effect into account in the weakly insulating regime~\cite{Altshuler85}. All of the raw data were taken at $T$=1.6\,K.} \label{Fig3}
\end{figure}

\begin{figure}
\centering
\includegraphics*[width=8.0 cm]{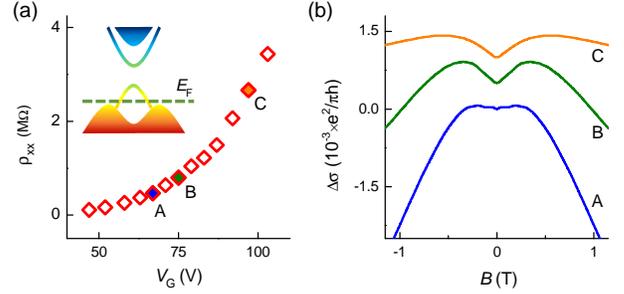}
\caption{(color online) (a) Gate voltage dependence of $\rho_\mathrm{xx}$ in a 2 QL (Bi$_{1-x}$Sb$_x$)$_2$Te$_3$ thin film. The inset shows a schematic band diagram. (b) MC curves of the same sample for three representative $\rho_\mathrm{xx}$ values, which are marked as points A-C in panel (a). The curves are vertically shifted for clarity. Data were taken at $T$\,=1.6 K.
} \label{Fig3}
\end{figure}

Such a change in the transport regime is also accompanied by a \emph{qualitative} change in magnetoconductivity (MC), which is defined as $\Delta\sigma(B)=\sigma(B)-\sigma(0)$. As shown in Fig.\,3, the films with weak disorder exhibit negative MC, which is a consequence of suppression of WAL by the magnetic field~\cite{Chen10,Checkelsky11,Chen11,Steinberg11,Kim12}. As disorder becomes stronger, the magnitude of the negative MC gradually decreases (Fig.\,3a-b), and eventually when the transport is in the strong localization ($g \ll 1$) regime, MC becomes positive in low magnetic fields (Fig.\,4).

For weakly disordered TI thin films ($g \gg 1$), the negative MC due to WAL can often be satisfactorily described with a simplified Hikami-Larkin-Nagaoka (HLN) equation:
\begin{equation}
\label{eq:WALtopo}
\Delta\sigma(B)\simeq
-\alpha\cdot\frac{e^2}{2\pi^2\hbar}
\left[
\psi\left(\frac{1}{2}+\frac{B_\phi}{B}\right)
-\ln\left(\frac{B_\phi}{B}\right)
\right],
\end{equation}
where $\psi(x)$ is the digamma function, $B_\phi=\hbar/(4De\tau_\phi)$ is the dephasing field, $D$ is the diffusion constant, and $\tau_\phi$ is the dephasing time. The prefactor $\alpha$ is equal to 1/2 and $-1$ for single channel transport in the WAL and the WL regimes, respectively. In TI thin films, however, multiple channels could be involved in the transport due to the existence of the top and bottom surface states, as well as possible conducting bulk.
Nevertheless, $\alpha \approx1/2$ has often been observed in TI thin films due to the coupling between various conducting channels~\cite{Chen10,Chen11,Checkelsky11,Steinberg11,KimY11,Taskin12,Kim13,Lang13,Brahlek14,Garate12}.


In the strong localization regime, it is generally accepted that MC is positive regardless of the strength of SOC \cite{Shapir89,Wohlman89,Schirmacher90,Meir91,Shklovskii91,Hernandez92,ZhangYZ92,Hsu95,Lin98}. The positive MC observed in this regime can be attributed to the quantum interference between different paths for forward scatterings in the variable range hopping transport~\cite{Shklovskii91,Lin98,Wohlman89,Schirmacher90}. It is different from the quantum corrections due to WAL (WL), in which the magnetic field suppresses the destructive (constructive) interference for backscattered paths~\cite{Bergmann84}. Several groups suggested theoretically that the MC in the variable range hopping regime is quadratic in sufficiently weak magnetic fields and becomes linear (or nearly linear) at stronger fields~\cite{Shklovskii91,Wohlman89,Schirmacher90}, until it reaches saturation at $H_\mathrm{sat}\approx(\phi_0/R_\mathrm{hop}\,^2)(R_\mathrm{hop}/\xi)^{1/2}$, where $\phi_0=h/e$ is the flux quantum and R$_{hop}$ is the hopping distance~\cite{ZhangYZ92}. As a result, the saturation field is expected to increase when the sample becomes more disordered. All of these features are qualitatively in agreement with the experimental data shown in Fig.\,4b. As to the decrease in the MC observed at higher magnetic fields, it may be related to the field induced modifications of electron wavefunctions localized around impurities or defects, similar to that observed previously in semiconductors in the variable range hopping regime~\cite{Shklovskii84}.

For an electron system with strong SOC, increasing the strength of disorder has been found previously to cause a change from the negative MC in the WAL regime to the positive MC in the strong localization regime. In an experiment with ultrathin metal films, Hsu and Valles observed that there is a well defined conductivity, $G_0=e^2/(\pi h)$ (or $g=1/\pi)$, separating the samples with negative MC and those with positive MC~\cite{Hsu95}. In the present work, however, we observed the change of sign of low field MC from negative to positive in the 2\,QL thick (Bi$_{1-x}$Sb$_x$)$_2$Te$_3$ film only when $\rho_\mathrm{xx}$ exceeds 0.5 M$\Omega$ (or $g<0.05\ll1$), as shown in Fig.\,4.

Also noteworthy is the intermediate regime between the diffusive and the hopping types of transport. This regime, with $\rho_\mathrm{xx}$ typically in the range of $\sim$10-100\,k$\Omega$, can be readily achieved with 2-5\,QL thick TI films. We found that the low field MC in the ultrathin films can be fitted fairly well with the HLN equation despite two orders of magnitude change in $\rho_\mathrm{xx}$, as illustrated in Fig.\,3a-b. Nevertheless, when $\rho_\mathrm{xx}$ is greater than 10\,k$\Omega$, $\alpha$ becomes to be smaller than the ordinary values (i.e.\,$\alpha\gtrsim1/2$) for WAL. The results from four highly tunable ultrathin films are summarized in Fig.\,3c, which shows that there is a clear correlation between $\alpha$ and $\rho_\mathrm{xx}$: Beyond the diffusive regime, increasing $\rho_\mathrm{xx}$ leads to greater suppression of WAL (i.e.\ smaller $\alpha$). When $\rho_\mathrm{xx}$ reaches 100\,k$\Omega$, $\alpha$ drops to nearly zero. It should be noted that $\alpha$ remains positive as long as $\rho_\mathrm{xx}$ is smaller than 100\,k$\Omega$.

Similar reductions in the $\alpha$ values previously reported for TI ultrathin films~\cite{Taskin12,Kim13,Lang13,KimY11} are also related to this intermediate disorder regime. An earlier study of 10\,QL thick Ca-doped Bi$_2$Se$_3$ thin films also found substantial reduction of $\alpha$ when $\rho_\mathrm{xx}$ is increased to $\sim$30\,k$\Omega$~\cite{Chen10}. It was pointed out by some of us~\cite{Chen11} and subsequently by Brahlek et al.~\cite{Brahlek14b} that the transport beyond the diffusive regime needs to be considered to account for observed suppression of WAL. Knowledge learned from conventional semiconductors~\cite{Minkov04,Lin98} and the data in Fig.\,3 are useful to gain further insight. For disorder stronger than that required for well-defined diffusive transport, quantum corrections for higher order terms of $1/g$ need to be taken into account. In the case of weak SOC, Minkov et al.~\cite{Minkov04} showed that the HLN equation could be valid, but the prefactor has to be reduced to $\alpha=[1-2/(\pi g)]$. Correspondingly, one would expect $\alpha=1/2-1/(\pi g)$ as an extension to the case of WAL in ultrathin TI films. Fig.\,3c shows that experimental $\alpha$ values roughly follow this trend when $g>1$. Further decreasing in conductivity (i.e. $g<1$) makes the perturbation treatment described in Ref.~\cite{Minkov04} no longer valid. Nevertheless, $\alpha$ values approaching zero obtained for $g\ll1$ are consistent with previous results on the traditional 2D electron systems with both weak~\cite{Minkov04} and strong~\cite{Lin98} SOC strengths. Such strong suppression of the negative MC can thus be interpreted as a result of the disorder-driven crossover from the so-called weakly insulating regime ($1<g<3$) toward the strong localization ($g\ll1$) regime. The positive MC and the variable range hopping transport related to the latter regime have not been observed perviously in TI materials.

Finally, we note that the suppression of WAL in ultrathin TI films may also take place as a consequence of a crossover from WAL to WL. Lu et al.~\cite{LuHZ11} suggested that, when chemical potential $\mu$ is comparable to the hybridization gap $\Delta$, a spin texture with a spin component perpendicular to the surface develops, and the Berry phase is modified to $\phi$=$\pi$(1-$\Delta$/2$|\mu|$). Therefore, tuning the chemical potential relative to the hybridization gap can in principle induce a crossover between WAL and WL. In this work, however, we were only able to observe the positive MC in the strong localization regime. We are also not aware of any previous work reporting observation of the positive MC with TI thin films in low magnetic fields while maintaining the electron transport in the \emph{weakly} disordered (diffusive) regime.  Therefore, significant improvement in the carrier mobility in ultrathin TI films is needed in order to unambiguously confirm the predicted crossover to WL due to the hybridization gap.

In summary, strong tuning of conductivity in the ultrathin TI films has allowed us to systematically study the transport properties from the WAL to the strong localization regime. Both temperature and disorder can drive such a crossover, which takes place when the phase coherence length exceeds the localization length. The suppression of negative MC and the appearance of positive MC in ultrathin TI films can be explained as a consequence of disorder-driven crossover between WAL and the strong localization. The topologically trivial insulating phases realized in ultrathin TI films~\cite{Linder09,LiuC10,Lu10} may provide a valuable platform for pursuing topological Anderson insulators, a novel state of matter also driven by disorder~\cite{LiJ09,JiangH09,Groth09}.

\begin{acknowledgements} \emph{Acknowledgements}: We thank H. Jiang, H. Z. Lu, A. D. Mirlin, S. Q. Shen and J. R. Shi for stimulating discussions. This work was supported by the National Basic Research Program (Project No: 2012CB921703, 2012CB921300 \& 2015CB921102), the National Science Foundation of China (Project Nos. 91121003, 11325421 \& 61425015), and the Chinese Academy of Sciences.
\end{acknowledgements}

\end{document}